\RequirePackage{amsthm}

\documentclass[pdflatex, sn-basic]{sn-jnl}

\usepackage{graphicx}%
\usepackage{multirow}%
\usepackage{amsmath,amssymb,amsfonts}%
\usepackage{amsthm}%
\usepackage{mathrsfs}%
\usepackage[title]{appendix}%
\usepackage{xcolor}%
\usepackage{textcomp}%
\usepackage{manyfoot}%
\usepackage{booktabs}%
\usepackage{algorithm}%
\usepackage{algorithmicx}%
\usepackage{algpseudocode}%
\usepackage{listings}%

\usepackage{tabularx}
\usepackage{float}
\usepackage{bm}
\usepackage{bbm}
\usepackage{mathtools}

\usepackage{dsfont}
\usepackage{caption}
\usepackage{subcaption}
\usepackage{colortbl}
\usepackage{multirow}
\usepackage{stmaryrd}
\usepackage[normalem]{ulem}

\usepackage{xspace}
\usepackage{comment}
\usepackage{thmtools, thm-restate}
\usepackage{verbatim}

\usepackage{array}
\newcolumntype{R}[1]{>{\raggedleft\arraybackslash}p{#1}}
\newcolumntype{L}[1]{>{\raggedright\arraybackslash}p{#1}}
\newcolumntype{C}[1]{>{\centering\arraybackslash}p{#1}}

\newcommand{\paperacronym}{\textrm{PBNeF}}

\usepackage[commandnameprefix=always, commentmarkup=uwave, defaultcolor=brown, draft]{changes}
\definechangesauthor[name=You]{EK}
\definechangesauthor[name=Todo]{todo}

\theoremstyle{thmstyleone}%
\theoremstyle{thmstyletwo}%

\theoremstyle{thmstylethree}%

\begin{document}

\title[End-to-End Reaction Field Energy]{End-to-End Reaction Field Energy Modeling via Deep Learning based Voxel-to-voxel Transform}


\author[1]{\fnm{Yongxian} \sur{Wu}}\email{yongxian.wu@uci.edu}

\author[1]{\fnm{Qiang} \sur{Zhu}}\email{qiangz11@uci.edu}

\author*[1]{\fnm{Ray} \sur{Luo}}\email{rluo@uci.edu}

\affil[1]{\orgdiv{Department of Chemical and Biomolecular Engineering, Molecular Biology and Biochemistry, Materials Science and Engineering, and Biomedical Engineering}, \orgname{University of California, Irvine}, \orgaddress{\city{Irvine}, \postcode{92697}, \state{California}, \country{United States}}}




\abstract{In computational biochemistry and biophysics, understanding the role of electrostatic interactions is crucial for elucidating the structure, dynamics, and function of biomolecules. The Poisson-Boltzmann (PB) equation is a foundational tool for modeling these interactions by describing the electrostatic potential in and around charged molecules. However, solving the PB equation presents significant computational challenges due to the complexity of biomolecular surfaces and the need to account for mobile ions. While traditional numerical methods for solving the PB equation are accurate, they are computationally expensive and scale poorly with increasing system size. To address these challenges, we introduce \paperacronym{}, a novel machine learning approach inspired by recent advancements in neural network-based partial differential equation solvers. Our method formulates the input and boundary electrostatic conditions of the PB equation into a learnable voxel representation, enabling the use of a neural field transformer to predict the PB solution and, subsequently, the reaction field potential energy. Extensive experiments demonstrate that \paperacronym{} achieves over a 100-fold speedup compared to traditional PB solvers, while maintaining accuracy comparable to the Generalized Born (GB) model.}




\maketitle

\section{Introduction}\label{sec:intro}
In the realm of computational biochemistry and biophysics, understanding the role of electrostatic interactions is paramount for elucidating the structure, dynamics, and function of biomolecules. Electrostatic forces are not only crucial for the stability and formation of biomolecular structures but also fundamentally influence biomolecular interactions such as protein-ligand binding,~\citep{sharp1990electrostatic,onufriev2013protonation}, enzyme catalysis~\citep{warshel1984calculations,page1971entropic,kraut2003challenges} and DNA transcription processes.~\citep{record1976ion,von1984protein,zakrzewska1996poisson} The Poisson-Boltzmann (PB) equation stands as a cornerstone in the modeling of these electrostatic interactions by describing the electrostatic potential in and around charged objects like proteins and DNA in an ionic solution.~\citep{fogolari2002poisson,baker2001electrostatics}

The importance of accurately and efficiently solving the Poisson-Boltzmann equation cannot be overstated. Accurate computation of PB electrostatic energies aids in a deeper understanding of the molecular mechanisms underpinning critical biological processes. For instance, in drug design, accurate electrostatic calculations can predict the strength and specificity of a drug molecule binding to its target protein, directly influencing the efficacy and safety profile of therapeutic interventions.~\citep{davis1990electrostatics} Furthermore, in protein engineering, precise electrostatic modeling helps in predicting how mutations affect protein stability, guiding the design of proteins with desired stability and activity.~\citep{warshel1976theoretical}

However, solving the PB equation poses significant computational challenges due to the complexity of biomolecular surfaces and the need to account for the mobile ionic environment. Traditional numerical methods for solving the PB equation, such as finite difference~\citep{nicholls1991rapid,bruccoleri1997finite,grant2001smooth} and finite element methods~\citep{holst2012adaptive,chen2007finite}, while accurate, are computationally intensive and scale poorly with system size. This scaling issue becomes a bottleneck in simulations involving large biomolecular complexes or when multiple simulations are required, such as in molecular dynamics studies or high-throughput screening scenarios.~\citep{lu2008recent} Therefore, the development of faster and equally accurate computational approaches for solving the PB equation is critical. Such advancements would not only democratize high-accuracy electrostatic calculations for broader sections of the scientific community but also enable real-time applications in computational biology and bioinformatics platforms. Improving the speed and accuracy of PB calculations opens the door to more dynamic and interactive modeling of biomolecular systems, potentially transforming our approach to understanding and manipulating biological systems at the molecular level.~\citep{baker2001electrostatics}

To address these challenges, we introduce \paperacronym{}, a novel machine learning approach inspired by recent advancements in neural network-based partial differential equation solvers, such as Fully Convolutional Network Solver~\citep{zhu2018bayesian}, Fourier Neural Operator~\citep{li2020fourier}, and Physics-informed Neural Solver~\citep{,pan2020physics}. Our method formulates the input and boundary electrostatic conditions of the PB equation into a learnable voxel representation, enabling the use of a neural field transformer to predict the PB solution and, subsequently, the reaction field potential energy. Extensive experiments demonstrate that \paperacronym{} achieves over a 100-fold speedup compared to traditional PB solvers, while maintaining accuracy comparable to the Generalized Born (GB) model.

In conclusion, the quest for efficient and accurate solutions to the Poisson-Boltzmann equation is more than a theoretical endeavor—it is a practical imperative that directly impacts the fields of molecular biology, pharmacology, and beyond. As we push the boundaries of what is computationally feasible, the insights gained from these electrostatic calculations will undoubtedly lead to groundbreaking advancements in science and medicine.

\section{Methods}\label{sec:methods}

\subsection{Definition of Poisson-Boltzmann Energy}
The electrostatic potential of biomolecular systems is an important property that has been widely utilized in the fields of drug design, protein-protein interactions, and enzyme catalysis~\citep{gilson1988calculating,wang2000well,baker2004poisson}. The Poisson-Boltzmann (PB) model, using continuum electrostatics to analyze, has garnered significant attention in both the mathematical and biophysical communities for its effectiveness in describing the electrostatic potential of biomolecular systems~\citep{honig1995classical,baker2001electrostatics}. To analyze the potential property of a biomolecular system in a typical implicit solvation framework (i.e., both solute and solvent molecules are represented as continua), the corresponding electrostatic potential described by PB model~\citep{nicholls1991rapid} can be formulated as:
\begin{equation}
  \nabla \cdot \left(\epsilon(\mathbf{r}) \nabla \phi(\mathbf{r})\right) = -\rho(\mathbf{r}),
  \label{eq:principle}
\end{equation}
where $\phi(\mathbf{r})$ is the electrostatic potential value defined at a field position $\mathbf{r}$, $\epsilon$ is the dielectric constant, $\phi$ is the electrostatic potential, and $\rho$ is the charge density. Despite its widespread use and success, the PB model presents a great challenge in directly solving the partial differential equation (PDE) in (\ref{eq:principle}), particularly when applied to large biomolecules. To improve the efficiency on applying the PB model to analyze electrostatic potentials, prior work~\citep{davis1991dielectric} is proposed to linearize the PB equation as follows:
\begin{equation}
  \nabla \cdot (\epsilon(\mathbf{r}) \nabla \phi(r)) - \lambda(\mathbf{r}) \sum_{i} n_i q_i^2 \phi(\mathbf{r})/kT = -\rho(\mathbf{r}),
\end{equation}
where $\lambda$ is the masking function for the Stern layer, $n_i$ is the number density of counterions, $q_i$ is the charge of the counterion, $k$ is the Boltzmann constant, and $T$ is the temperature. Many PDE solving techniques, such as finite difference method (FDM)~\citep{davis1989solving,holst1993multigrid,rocchia2001extending,luo2002accelerated} and preconditioned conjugate gradient methods~\citep{baker2001electrostatics}, have been applied to boost the solving process to calculate the electrostatic potential value $\phi(\mathbf{r})$. The energy of linear Poisson-Boltzmann model (EPB) defined on the calculated electrostatic potential value $\phi(\mathbf{r})$ can be formulated as:
\begin{equation}
  \Delta G = \frac{1}{2} \sum_{i}^{N_a} q_i \phi(\mathbf{r}_i)
\end{equation}
where \(q_i\) is the charge of atom \(i\) and \(\phi(\mathbf{r}_i)\) is the reaction field potential value of atom \(i\) at a field position $\mathbf{r}_i$. Although previous works have achieved a great improvement in speeding up the biomolecular electrostatics calculation, the computationally intensive characteristic of explicitly solving PB model remains a strong constraint on scaling up to large biomolecules especially with tens of thousands to millions of atoms.

\subsection{\paperacronym{}: Implicit EPB Solver by Neural Field Transform}
Recently, neural networks have been shown to be efficient on running simulation, such as molecule dynamics~\citep{zhang2018deep,batzner20223}, computational fluid dynamics~\citep{thuerey2021physics,raissi2020hidden}, and solving partial differential equations~\citep{raissi2019physics,sirignano2018dgm,li2020fourier}, due to its impressive parallel inference ability on accelerators. Inspired by the great success achieved by neural networks, we propose an implicit PDE solver framework for the poisson-boltzmann model by carefully parameterize the neural network without needing of explicitly solving the partial differential equation in the PB model.

Our model, named \paperacronym{}, takes as input a voxelized field representation of the molecular and outputs the predicted electrostatic potential value $\phi(\mathbf{r})$ at each field position. The \paperacronym{} model consists of two major components: a learnable neural field voxelized representation (VNeFR), and a neural field transformer (NeFT). The neural field transformer is based on the learnable neural field representation aiming to learn the spatial interactions between field positions in the molecular structure. This component captures the intricate spatial relationships and dependencies within the neural field representation, enabling the model to understand the complex electrostatic environment.

\subsubsection{Learnable Neural Field Voxelized Representation}
Motivated by the point cloud representation of molecular structures, we propose a novel deep learning approach to model the end-to-end reaction field energy calculation. To preserve the structure property of the molecule, we represent it in the voxel format. Within this representation, the entire molecule space is discretized into a three-dimensional (3D) voxel grid with the size of $N_D\times N_H \times N_W$, which is denoted as $\mathcal{G} = \{\mathbf{g}^{(1)}, \ldots, \mathbf{g}^{(N_g)}\}$ containing $N_g$ grid points. For each grid point $\mathbf{g}^{(i)}$, $\mathbf{g}^{(i)} \in [N_D]\times [N_H] \times [N_W]$ signifies the coordinate of the $i$th grid point in the 3D voxel grid. The square brackets $[N_D]$ indicate the range or collection of integers from $0$ up to $N_D-1$, which is also applied for $N_H$ and $N_W$. The distance between two nearest neighbor grid points is a constant value $s \in (0, 1]$ (also referred to as the grid spacing). The concept of point cloud was then introduced here for the representation of 3D molecular coordinates, which is a set of points $\mathcal{P} = \{\boldsymbol{p}^{(1)}, \ldots, \boldsymbol{p}^{(N_a)}\}$,~\citep{ladicky2017point,qi2017pointnet,achlioptas2018learning,wu2023grid} where $\boldsymbol{p}^{(i)} = (x^{(i)}, y^{(i)}, z^{(i)}) \in \mathbb{R}^3$ represents the atomic coordinates of atom $i$ in the 3D space, $N_a$ is the total atom number of a molecular. The atomic coordinates are then projected onto the grid space $\mathcal{G}$, where it consists of a set of data points with each point denoting the specific position of atoms in molecular systems. Therefore, for a molecule containing $N$ atoms, it can be expressed as $\mathcal{M} = \{\boldsymbol{a}^{(1)}, \ldots, \boldsymbol{a}^{(N)}\}$, where $\boldsymbol{a}^{(i)} = (x^{(i)}, y^{(i)}, z^{(i)}) \in [N_D]\times [N_H] \times [N_W]$ represents the atomic coordinates of atom $i$ in the voxel grid. In the context that follows, all features are discussed within the voxel grid space for the sake of brevity.

\begin{figure}[!htbp]
  \centering
  \begin{subfigure}[b]{0.4\textwidth}
    \centering
    \includegraphics[height=0.6\textwidth]{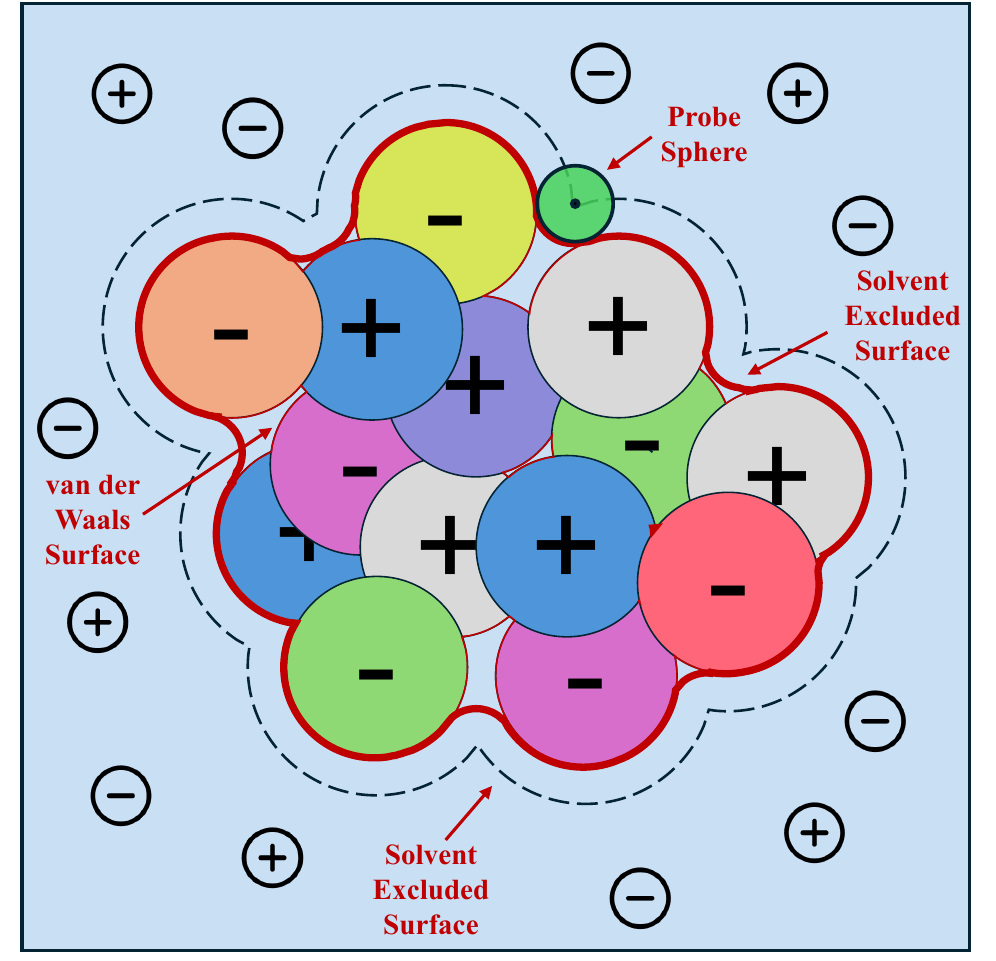}
    \caption{Atom charge and level-set value}
    \label{fig:example_feature_epb}
  \end{subfigure}
  \begin{subfigure}[b]{0.55\textwidth}
    \centering
    \includegraphics[height=0.4\textwidth]{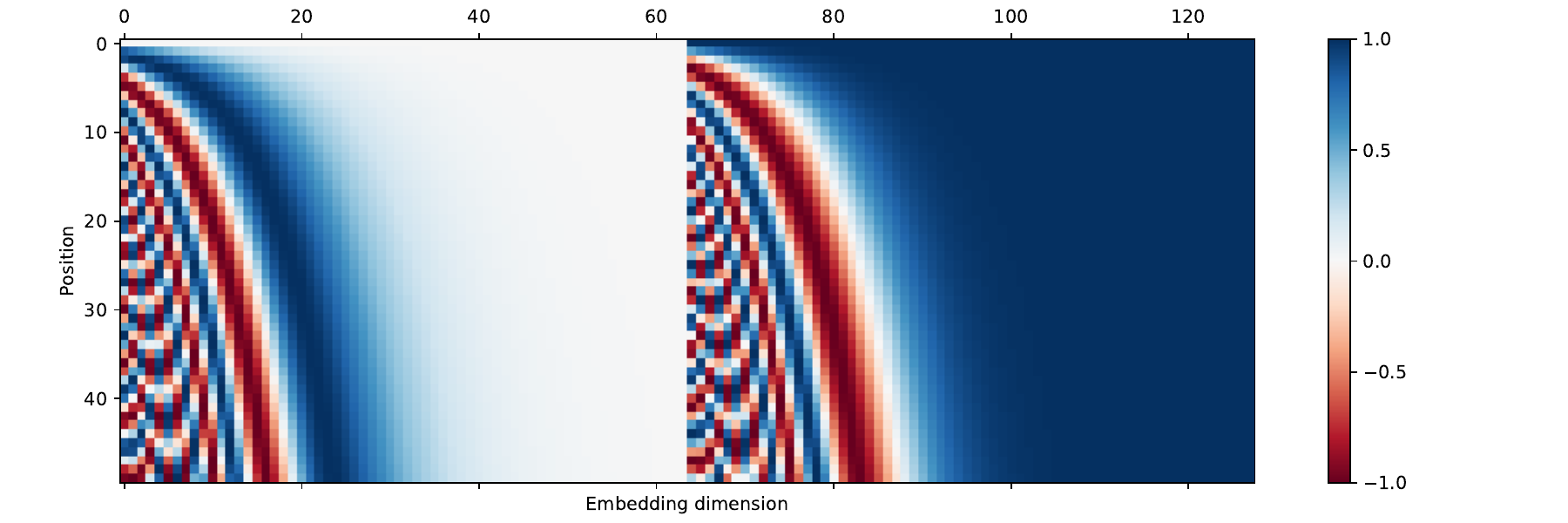}
    \caption{Example of high-frequency Fourier embedding}
    \label{fig:example_sinusoidal}
  \end{subfigure}
  \caption{
    \textbf{Illustration of molecular features and high-frequency Fourier embedding.} 
    (a) Representation of atom charge and Solvent Excluded Surface (SES). Different colors correspond to different types of atoms, with charges indicated by ``+'' and ``-'' symbols.
    (b) Example of high-frequency Fourier embedding applied to scalar features like atom charge $q^{(i)}$.
    }
  \label{fig:positional_encoding}
\end{figure}

According to the Poisson-Boltzmann model~\citep{holst1994poisson}, the EPB is defined by dielectric constant, electrostatic potential, and charge density, which are derived from atom charge, atom position, and level-set values. Instead of defining an explicit analytical partial differentiation equation, we aim to build an implicit neural network to learn the mapping from the molecular propertied to the EPB value. An illustration of molecular features is shown in Figure~\ref{fig:example_feature_epb}. For atom $i$, let's denote its atom charge as $q^{(i)}$. As the dielectric constant depends on the boundary of the molecule, the level-set value is utilized to indicate the molecule surface position. The level-set function has been extensively used for representing the SES of a molecule due to its convenience.~\citep{ye2010revised,wang2009achieving,wei2021machine,wu2023grid} The sign of level-set values $\ell^{(i)}$ indicates whether a grid point $\mathbf{g}_i$ is positioned outside-of-boundary or inside-of-boundary.~\citep{osher2003level} By considering the information above, we aim to enable the neural network be able to learn to solve the EPB value $\Delta G$ by learning an implicit partial differentiation equation solver.

\begin{figure*}
  \centering
  \includegraphics[width=0.9\textwidth]{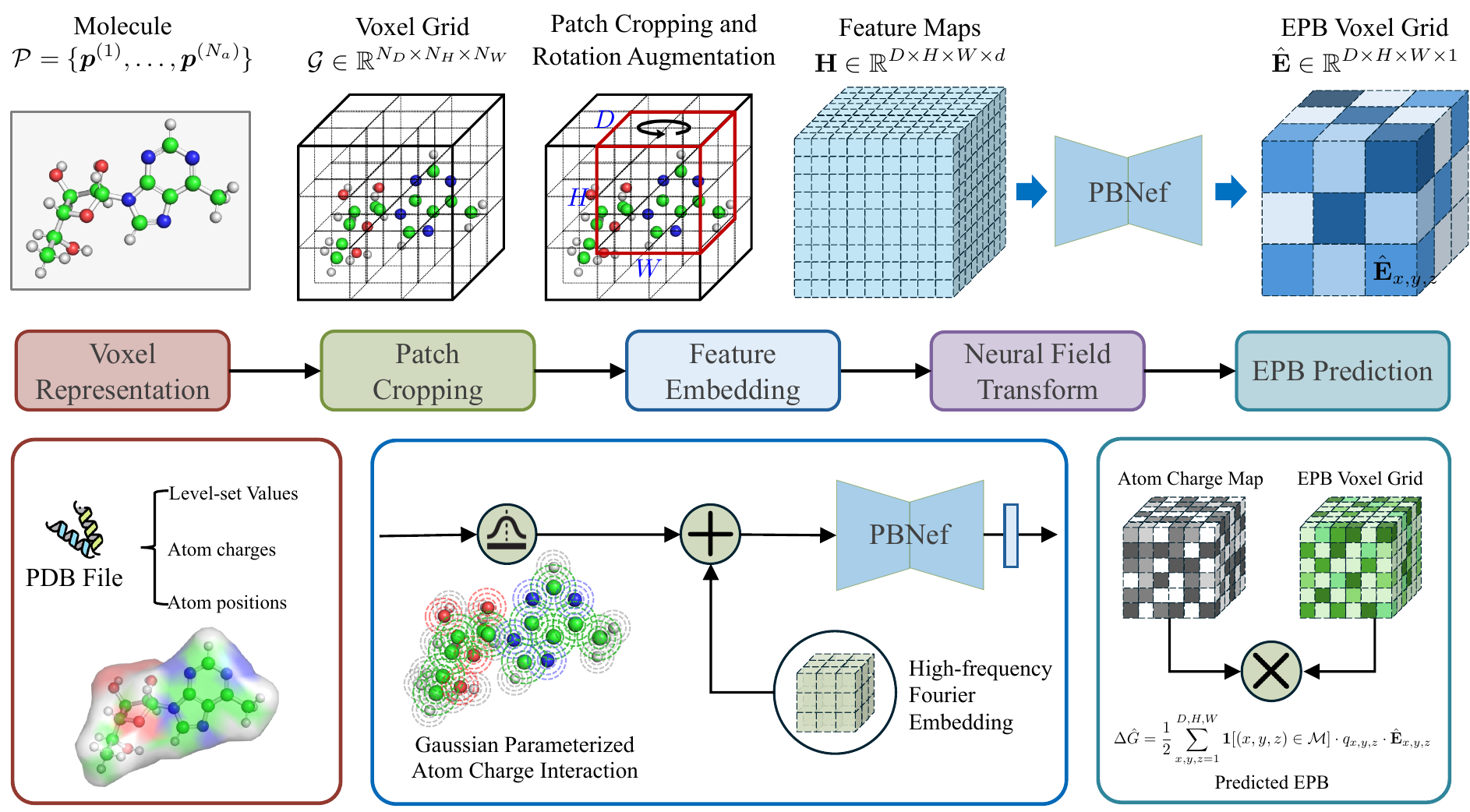}
  \caption{
        \textbf{The overview of the proposed PBNef method.} 
        The figure illustrates the workflow of the PBNef model for predicting electrostatic potential of biomolecules (EPB). The process begins with the voxel representation of a molecule, where atomic properties such as level-set values and charges are encoded into a 3D voxel grid. To address challenges in learning low-rank scalar features and capturing charge interactions, Fourier feature mapping is used to encode these features into high-frequency vectors. A Gaussian parameterized atom charge interaction model is employed to represent the interactions between atoms. The spatial interactions are understood through a neural field transformer that processes the 3D structure of the molecule. Given the potentially large size of voxel grids, a patch cropping strategy with random cropping and rotation augmentation is applied to improve model robustness. The final step involves the neural network predicting EPB values across the voxel grid, with results integrated into an atom charge map for EPB prediction. The lower portion of the figure highlights the key components, including the Gaussian parameterized interaction and high-frequency Fourier embedding, which enhance the model’s accuracy and robustness.
    }
  \label{fig:dlpb_overview}
\end{figure*}

However, there are some challenges to achieve this goal. First, the low-rank scalar features, level-set value $\ell^{(i)}$ and atom charge $q^{(i)}$ has been proved to be difficult to be learned directly by the neural network due to the limited semantics implicated in the scalar value.\citep{tancik2020fourier} To address this challenge, we propose the Fourier feature mapping strategy to encode these low-rank features into a high frequency vector, which enable the neural network to achieve a better performance. Second, to model charge interactions between any two atoms, the neural network should be able to capture the charge interactions. This motivates us to propose a learnable Gaussian distribution, which parameterizes a Gaussian blurry kernel to consider this type of interaction. To consider the spatial interaction, we propose a neural field transformer to understand the 3D structure of atoms in the voxel grid. Third, since the size of voxel grid of molecule can be tremendously large, the neural network should be able to predict the EPB value over molecules represented at any voxel size. Therefore, we propose a patch cropping strategy with random cropping and rotation data augmentation over the 3D voxel, which turns out to be effective in improving the model performance and robustness. Figure~\ref{fig:dlpb_overview} illustrates the overview of the proposed \paperacronym{} method. The details of these strategies are introduced in the following sections.

\paragraph{Patch Cropping}
The size of the voxel grid of a molecule can be tremendously large (i.e., the voxel size $N_D \times N_H \times N_W$ is too large to be computationally feasible due to the CPU memory constraint), which poses a challenge for the neural network to predict the EPB value over molecules represented at any grid size $s$. To address this challenge, we propose a patch cropping strategy over the 3D voxel grid, which enables \paperacronym{} to seamlessly do inference on voxel grids with any sizes. In general, a subregion of the voxel grid (i.e., a patch) is randomly selected to act as input to the neural network during the training stage. Consequently, a subregion of voxel grid with a patch size of $D \times H \times W$ is cropped from the original voxel grid. In the following content, we only consider features fall inside a cropped patch for brevity.

\paragraph{High-Frequency Fourier Embedding for Continuous Scalar Features} According to the neural tangent kernel (NTK) literature, a standard multi-layer perceptron (MLP) fails to learn high frequencies both in theory and in practice~\citep{tancik2020fourier}. However, in our EPB prediction problem, features are encoded in a low dimensional space but with high frequency difference, such as $q^{(i)}$ and $\ell^{(i)}$. To overcome this problem, we propose to use the High-Frequency Fourier Embedding to encode the low-rank scalar features into high-frequency vectors, which enables the neural network to capture the significant difference encoded in the scalar feature. For brevity, we describe the high-frequency Fourier embedding process for atom charge $q^{(i)}$ as an example. The embedding process for level-set value $\ell^{(i)}$ is similar.

The high-frequency Fourier embedding process can be denoted as an embedding function $\boldsymbol{e}_f: \mathbb{R} \rightarrow \mathbb{R}^d$, which takes a scalar feature as input and outputs a vector with $d$ dimensions. The formulation of $\boldsymbol{e}_f$ is defined as follows:
\begin{align}
  \boldsymbol{e}_f(q^{(i)})_j :  =
  \begin{cases}
    \sin(\omega_j \cdot \alpha \cdot q^{(i)} ), & \text{if } j = 2k     \\
    \cos(\omega_j \cdot \alpha \cdot q^{(i)} ), & \text{if } j = 2k + 1
  \end{cases},
  ~~\omega_j                      = \exp(-\frac{i}{d-1}\log\theta)
\end{align}
where $\boldsymbol{e}_f(q^{(i)}) \in \mathbb{R}^d$ is the encoded $d$-dimensional high-frequency embedding of the atom charge $q^{(i)}$ that the $j$th dimension represents the scalar feature encoded in the specific frequency $\omega_j$,  $\alpha$ is a scaling factor to normalize the charge value $q^{(i)}$ into the same scale with other scaler features (i.e., the level set value $\ell^{(i)}$), $\theta$ is a hyperparameter to control the frequency of the embedding vector, and $d$ is the dimension of the embedding vector. An example of high-frequency Fourier embedding is illustrated in Figure~\ref{fig:example_sinusoidal}.

Given the high-frequency encoding vector $\boldsymbol{e}_f(q^{(i)})$, we apply a multi-layer perceptron $\textrm{MLP}: \mathbb{R}^d \rightarrow \mathbb{R}^d$ to learn the feature hidden in different frequencies:
\begin{align}
  \boldsymbol{q}^{(i)} = \textrm{MLP}(\boldsymbol{e}_f(q^{(i)}))
\end{align}
where $\boldsymbol{q}^{(i)} \in \mathbb{R}^d$ is the learned feature representation of the atom charge $q^{(i)}$, $\textrm{MLP}(\cdot)$ is a two layer feedforward neural network with GeLU activation function~\citep{hendrycks2016gaussian}. With the learned embeddings for each atom $\{\boldsymbol{q}^{(i)}\}_{i=1}^{N_a}$, we scatter them into the voxel grid according to their atomic coordinates, which results the voxel-level feature representation of atom charge $\mathbf{H}^q \in \mathbb{R}^{D\times H \times W \times d}$. The feature representation of level-set value $\ell^{(i)}$ is encoded in the same way as the atom charge $q^{(i)}$ and results in the voxel-level feature representation of level-set value $\mathbf{H}^\ell \in \mathbb{R}^{D\times H \times W \times d}$.

\begin{figure}
  \centering
  \includegraphics[width=0.9\textwidth]{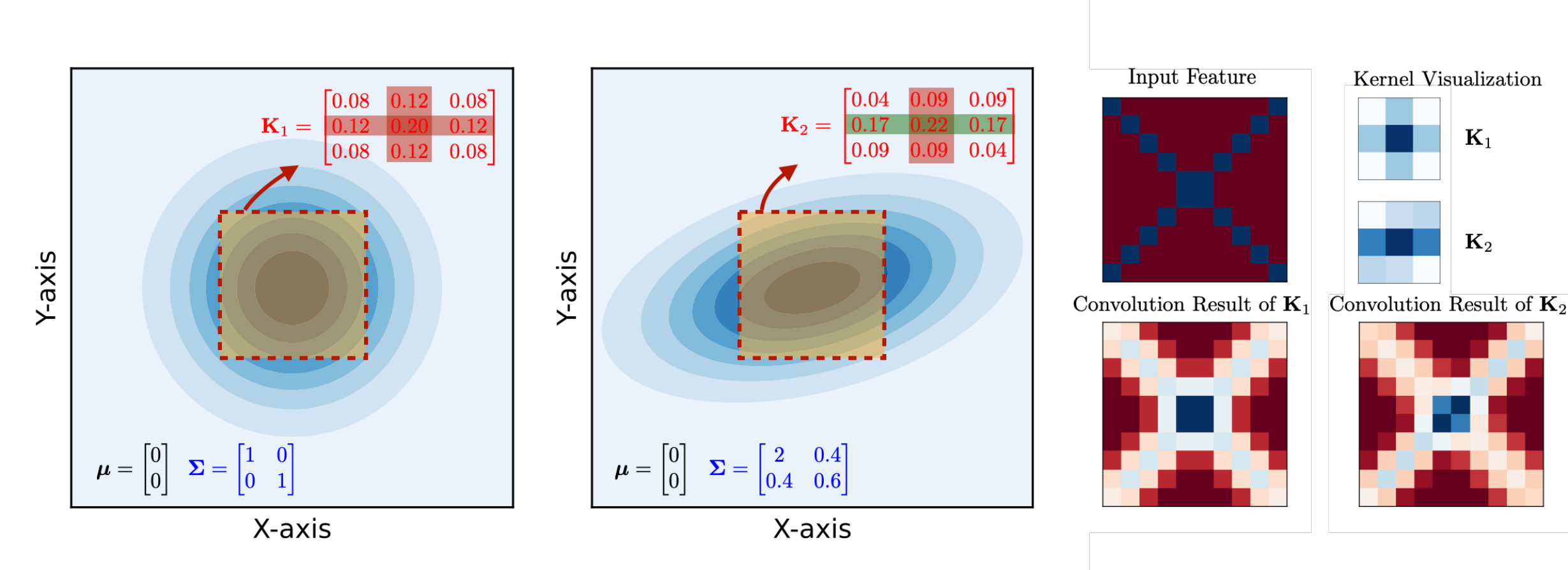}
  \caption[Gaussian Kernel for Atom Charge Distribution]{\textbf{Visualization of Gaussian kernels for atom charge distribution.} The figure illustrates the concept of using learnable Gaussian blur kernels, parameterized by a multivariate Gaussian distribution, to diffuse atom charge interactions to neighboring grid points in the molecular structure. The left panel shows a Gaussian kernel with a mean vector $\boldsymbol{\mu} = [0, 0]$ and a covariance matrix $\boldsymbol{\Sigma} = \begin{bmatrix} 1 & 0 \\ 0 & 1 \end{bmatrix}$, resulting in an isotropic distribution. The right panel demonstrates a kernel with the same mean vector but a different covariance matrix $\boldsymbol{\Sigma} = \begin{bmatrix} 2 & 0.4 \\ 0.4 & 0.6 \end{bmatrix}$, leading to an anisotropic distribution.}
  \label{fig:gaussian_kernel}
\end{figure}

\paragraph{Gaussian Parameterized Atom Charge Interaction} The atom charge distribution is a critical feature for capturing the electrostatic interactions between atoms in the molecular structure. To model the interaction among different atom charges, we propose a learnable Gaussian distribution that parameterizes a Gaussian blur kernel to distribute the atom charge to the nearby grids:
\begin{align}
  \tilde{\mathbf{H}}^{q}_{x,y,z} = \sum_{(i,j,k)\in \textrm{Neigh}(x,y,z)} \mathbf{H}^{q}_{i, j, k}\cdot \mathbf{K}(i - x, j - y, k-z), ~~ \mathbf{K} \sim \mathcal{N}(\boldsymbol{\mu}, \boldsymbol{\Sigma} \cdot s^2)
\end{align}
where $\tilde{\mathbf{H}}^{q}_{x,y,z}$ is the diffused atom charge feature at the grid point $(x,y,z)$, $\textrm{Neigh}(x,y,z)$ is the set of neighboring grid points of $(x,y,z)$, $\mathbf{K}$ is the Gaussian blur kernel to model the interaction between the atom charge at the center grid point with atom charges in its neighboring grid points. The kernel $\mathbf{K}$ is parameterized by a multidimensional Gaussian distribution $\mathcal{N}(\boldsymbol{\mu}, \boldsymbol{\Sigma} \cdot s^2)$, where $\boldsymbol{\mu} \in \mathbb{R}^3$ is the mean vector, $s$ is the grid space value, and $\boldsymbol{\Sigma}  \in \mathbb{R}^{3\times 3}$ is the covariance matrix controlling the strength and field of atom charge interactions. The reason of including the grid space value $s$ is the interaction relationship is defined on the atom distance of 3D structure instead of the distance of grid. Therefore, the interaction blur kernel should be scaled by the grid space value to ensure the interaction relationship is consistent with the atom distance:
\begin{align}
  \mathbf{K}(i - x, j - y, k-z) \Leftrightarrow \mathbf{K}((i - x)\cdot s, (j - y) \cdot s, (k-z) \cdot s)  ~~ \mathbf{K} \sim \mathcal{N}(\boldsymbol{\mu}, \boldsymbol{\Sigma}).
\end{align}
Instead of setting the pre-defined Gaussian kernel, we propose to learn the Gaussian kernel parameter $\boldsymbol{\Sigma}$ by the training process. The diagonal entries of $\boldsymbol{\Sigma}$ is trained together with model parameters to better consider the interaction relationship between atom charges in the molecular structure. In practice, the mean vector $\boldsymbol{\mu}$ and off-diagonal entries of $\boldsymbol{\Sigma}$ are fixed to zeros to ensure the interaction kernel is isotropic. The diffused atom charge feature $\tilde{\mathbf{H}}^{q}$ is then added with the original atom charge feature $\mathbf{H}^q$, and the level-set value feature $\mathbf{H}^\ell$ to form the final molecule representation:
\begin{align}
  \mathbf{H} = \mathbf{H}^q + \tilde{\mathbf{H}}^{q} + \mathbf{H}^\ell
\end{align}
where $\mathbf{H} \in \mathbb{R}^{D\times H \times W \times d}$ is the voxel-level feature representation of molecule. An example of Gaussian kernel for atom charge is illustrated in Figure~\ref{fig:gaussian_kernel}.

\subsubsection{Neural Field Transformer}
Given the voxel-level feature representation of molecule $\mathbf{H}$, we aim to learn the spatial interactions between voxels in the molecular structure. To achieve this goal, we propose a neural field transformer to understand the 3D structure of atoms in the voxel grid. In details, the neural field transformer $\mathcal{F}: \mathbb{R}^{D\times H \times W \times d} \rightarrow \mathbb{R}^{D\times H \times W \times 1}$ is a neural operator that takes a subregion of voxel feature as input learning the spatial dependencies in the molecular structure and outputs a transformed voxel with predicted EPB values at each grid point
\begin{align}
  \hat{\mathbf{E}} & = \mathcal{F}(\mathbf{H})
\end{align}
where $\hat{\mathbf{E}} \in \mathbb{R}^{D\times H \times W \times 1}$ is the predicted energy at each grid point.

\begin{figure}[t]
  \centering
  \includegraphics[width=0.5\textwidth]{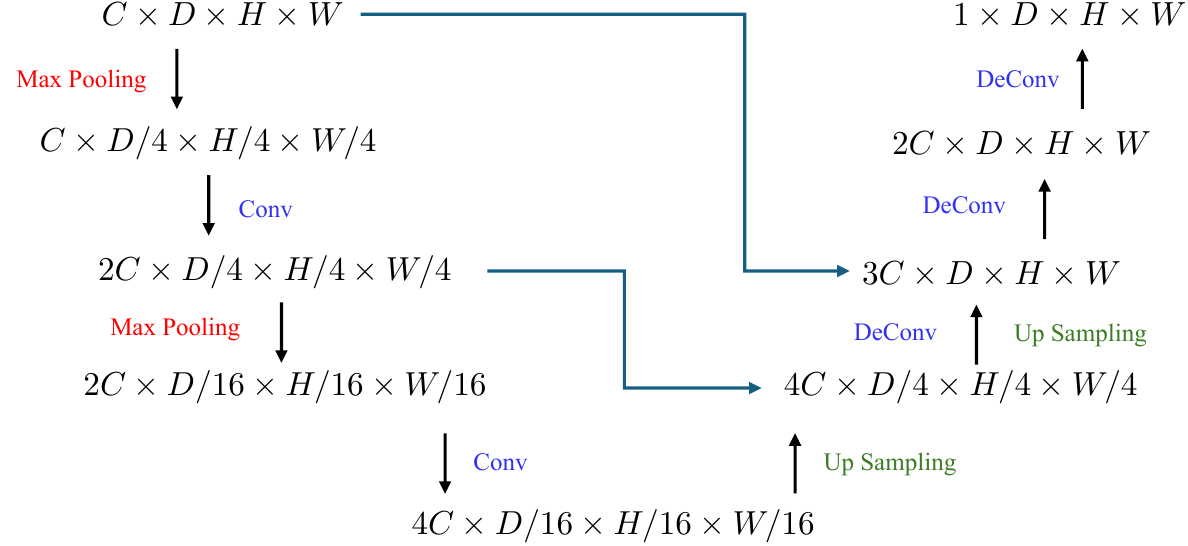}
  \caption{Illustration of Neural Field Transformer using U-Net like structure.}
  \label{fig:3d-unet}
\end{figure}

Inspired by~\citet{li2020fourier}, we propose two different backbones to implement the neural field transformer, 1) U-Net like structure and 2) Fourier neural operator. Figure~\ref{fig:3d-unet} illustrates an example of neural field transforer using a U-Net like structure. The neural field transformer consists of an encoder-decoder architecture with skip connections to capture the spatial interactions between voxels at different scales. The encoder consists of multiple convolutional layers with GeLU activation functions to extract the spatial features of the input voxel. The decoder consists of multiple deconvolutional layers with GeLU activation functions to upsample the spatial features and generate the final voxel-level prediction. Skip connections are added between the encoder and decoder to preserve the spatial information and improve the model's performance. The output of the neural field transformer is a voxel $\hat{\mathbf{E}} \in \mathbb{R}^{D\times H \times W \times 1}$ with predicted EPB values at each grid point, which can be used to calculate the reaction field energy of the molecular structure. The fourier neural operator can be applied to the voxel input in a similar way, which utilizes fourier transform to capture the global atom interaction in the space. Consequently, the total EPB value of the input voxel is calculated as the sum of the EPB values at each grid point:
\begin{align}
  \Delta \hat{G}  = \frac{1}{2}\sum_{x,y,z=1}^{D,H,W} \mathbf{1}[(x,y,z) \in \mathcal{M}] \cdot  q_{x,y,z} \cdot \hat{\mathbf{E}}_{x,y,z}
\end{align}
where $\Delta \hat{G}$ is the predicted EPB energy of the grid subregion, $\mathbf{1}[\cdot]$ is the indicator function that returns 1 if a grid point $(x,y,z)$ contains an atom belong to $\mathcal{M}$ and 0 otherwise. The EPB value at each grid point $\hat{\mathbf{E}}_{x,y,z}$ is multiplied by the according atom charge $q_{x,y,z}$ and the indicator function to calculate the total EPB value of the grid subregion. As a result, a mean-squared error (MSE) loss function is used to train the neural field transformer:
\begin{align}
  \mathcal{L}_{MSE} = \frac{1}{N} \sum_{i=1}^{N} (\Delta \hat{G}^{(i)} - \Delta G^{(i)})^2
  \label{eq:dlpb_obj}
\end{align}
where $N$ is the number of training patches. Since the neural field transformer is fully differentiable, the loss value can be optimized by standard backpropagation algorithm.

\paragraph{Random Cropping and Rotation Data Augmentation} During the training stage, the neural field transformer is trained on a subregion of the voxel grid with a patch size of $D \times H \times W$. To improve the model's performance and robustness, we propose a random cropping and rotation data augmentation strategy to increase the diversity of the training data. In this strategy, a subregion of the voxel grid is randomly selected as the input to the neural field transformer. The subregion is then rotated by a random angle around the any one axis to augment the training data. This data augmentation strategy enables the neural network to learn the spatial interactions between voxels at different orientations, improving the model's generalization ability and robustness.



\section{Results and Discussions}

\subsection{Experimental Setup}
\paragraph{Benchmark Dataset Preparation}
In order to train a model that can handle all types of biomolecular structures, we used all biomolecules from different systems in the AMBER~\citep{case2023ambertools} PBSA benchmark suite. This dataset contains a wide range of different structure biomolecular protein, nucleic acid, and protein complex structures, where the structural flexibility and pattern are quite different between each other. There are 574 biomolecular protein structures in the dataset having atomic counts ranging from 377 to 8,254, providing a diverse set of geometries. There are 365 nucleic acid structures in the dataset having various sizes with the number of atoms ranging from 250 to 5,569, and exhibit diverse geometries. There are 554 protein complex structures in the dataset demonstrating a wide range of geometries, with atom counts varying from 660 to 142,324. We generated the training data using a custom PBSA program. This program utilized default atomic cavity radii, extracted from topology files, and a solvent probe radius set to 1.4 \AA. Data collection for benchmarking was performed at grid spacings of 0.35 \AA, 0.55 and, 0.75 \AA, covering the entire dataset. All other parameters were maintained at their default values as specified in the PBSA module of the AMBER 23 package.

Through this meticulous process, we acquired a dataset comprising 1493 biomolecules in total. To stratify the dataset, 20\% of the data was allocated for testing purposes, while the remaining 80\% was further divided into training and evaluation datasets. We used a test size of 0.1 and a consistent random seed for reliable reproducibility during the train-test split.

\paragraph{Baselines} To evaluate the performance of \paperacronym{}, we conducted a series of experiments focusing on both accuracy and speed. Given the importance of balancing accuracy with computational efficiency, particularly in dynamic applications, we compared our model’s performance with the Generalized Born (GB) model implemented in AMBER and the Poisson-Boltzmann (PB) model implemented in AMBER PBSA, a widely used approximation for calculating the electrostatic component of solvation free energy~\citep{dominy1999development, bashford2000generalized,calimet2001protein,tsui2000molecular,wang2003implicit,gallicchio2004agbnp,nymeyer2003simulation}. Additionally, we trained a variant of our model called \paperacronym{}-Lite, which does not require the level-set feature to predict EPB values. In our initial experiment, we found that both U-Net like structure and Fourier neural operator achieve a similar performance. 

\subsection{Evaluation Metrics}

The coefficient of determination ($\mathrm{R}^2$) and the Mean Absolute Error (MAE) are two commonly utilized performance metrics.

\paragraph{Coefficient of Determination}

The coefficient of determination, also known as $\mathrm{R}^2$, quantifies the proportion of the variance in the dependent variable that can be predicted from the independent variables~\citep{nagelkerke1991note}. In our setting, the $\mathrm{R}^2$ score can be computed as follows:
\begin{equation}
  \mathrm{R}^2 = 1 - \frac{\sum_{i=1}^{n} (\Delta G^{(i)} - \Delta\hat G^{(i)})^2}{\sum_{i=1}^{n} (\Delta G^{(i)} - \Delta{\bar G})^2}
\end{equation}
where $n$ is the number of samples, $\Delta G^{(i)}$ is the truth level-set value that directly extracted from the AMBER PBSA benchmark suite, $\Delta\hat G^{(i)}$ is the predicted value by our model, and $\Delta{\bar G}$ is the mean of the observed values. An $\mathrm{R}^2$ score of 1 indicates perfect prediction, while a score of 0 suggests that the model does not improve predictions over using only the mean of the target variable.

\paragraph{Mean Absolute Error}
The Mean Absolute Error (MAE) is another popular metric used in model evaluation. It is defined as the average of the absolute differences between the predicted and observed values~\citep{willmott2005advantages}. In our setting, the MAE can be calculated as:
\begin{equation}
  \mathrm{MAE} = \frac{1}{n} \sum_{i=1}^{n} |\Delta\hat G^{(i)} - \Delta G^{(i)}|
\end{equation}
where $n$ is the number of samples, $\Delta G^{(i)}$ is the truth level-set value that directly extracted from the AMBER PBSA benchmark suite, and $\Delta\hat G^{(i)}$ is the predicted value by our model. MAE offers a straightforward interpretation, representing the average magnitude of the errors made by the model in its predictions, regardless of their direction. Lower MAE values indicate better predictive accuracy, with an MAE of 0 signifying a perfect predictor.


\subsection{Model Accuracy Analysis}

\begin{figure}[!htbp]
  \centering
  \begin{subfigure}[b]{0.45\textwidth}
      \includegraphics[width=0.9\textwidth]{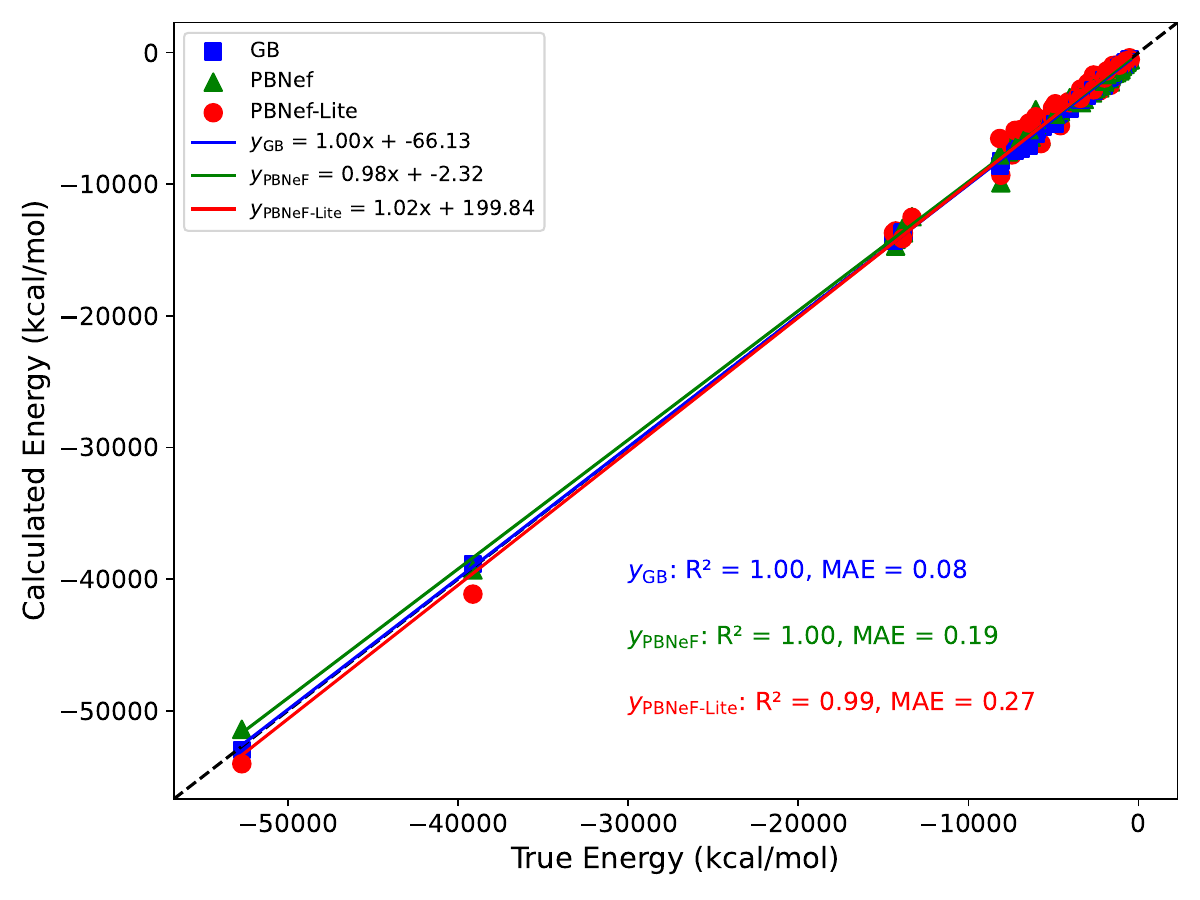}
      \caption{EPB accuracy comparison}
      \label{fig:energy_all_system}
  \end{subfigure}
  \begin{subfigure}[b]{0.45\textwidth}
      \includegraphics[width=0.9\textwidth]{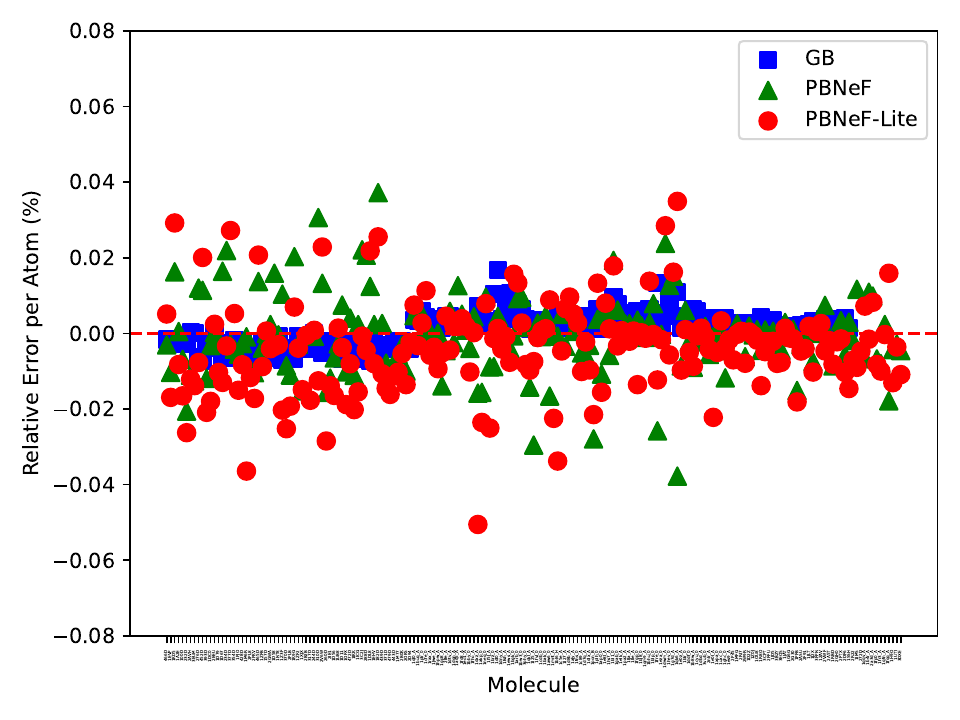}
      \caption{Relative error of predicted EPB value}
      \label{fig:relative_error_all_system}
  \end{subfigure}
  \caption[Comparison of EPB Model Accuracy and Relative Error Analysis]{\textbf{(a) EPB Accuracy Comparison:} Scatter plot showing the relationship between predicted and true energy values for the GB model (blue squares), \paperacronym{} (green triangles), and \paperacronym{}-Lite (red circles). The linear regression lines and corresponding equations illustrate the accuracy of each model, with $R^2$ and MAE values providing a quantitative measure of their performance. \textbf{(b) Relative Error of Predicted EPB Values:} Plot displaying the relative error per atom for each molecule across the three models, highlighting the consistency and variability in their predictions.}
  \label{fig:pb_acc}
\end{figure}

To assess the accuracy of our \paperacronym{} model, we compared its predictions against the ground truth values obtained from the AMBER PBSA benchmark suite and the energy calculated by the generalized Born (GB) solvation model. The results are shown in Figure~\ref{fig:pb_acc}. We used two common metrics, the coefficient of determination ($\mathrm{R}^2$) and the Mean Absolute Error (MAE), to evaluate the model’s performance.

Figure~\ref{fig:energy_all_system} provides a direct comparison of the calculated energies to the true energies across the GB, \paperacronym{}, and \paperacronym{}-Lite models. The $\mathrm{R}^2$ score quantifies the proportion of the variance in the dependent variable that can be predicted from the independent variables. Three sets of data points are shown: one represented by blue squares for the GB model, another by green triangles for \paperacronym{}, and the third by red circles for the \paperacronym{}-Lite variant. The linear regression lines for all methods are also included in the plot. The blue line represents the linear fit for the GB model with an equation of $y_{GB} = 1.00x - 66.13$, demonstrating a slope very close to 1, indicating an almost perfect proportional relationship, but with a slight offset. The green line represents the linear fit for $y_{\paperacronym{}}$ with an equation of $y_{\paperacronym{}} = 0.98x - 2.32$, showing a slope close to 1 and a minimal offset, indicating a high level of accuracy in the calculated energies using this method as well. Additionally, the red line for $y_{\paperacronym{}-Lite}$ has an equation of $y_{\paperacronym{}-Lite} = 1.02x + 199.84$, with a slope slightly above 1, indicating a small overestimation in the predicted energies.

The $R^2$ values for the GB and \paperacronym{} models are both 1.00, indicating that these models explain all the variability of the response data around their means, which suggests a perfect fit. For \paperacronym{}-Lite, the $R^2$ value is 0.99, still indicating a strong fit but with slightly more variability. The mean absolute error (MAE) values are 0.08 for the GB model, 0.19 for \paperacronym{}, and 0.27 for \paperacronym{}-Lite, indicating that the GB model has the lowest errors, followed by \paperacronym{}, and finally, \paperacronym{}-Lite with the highest errors. Overall, this figure demonstrates that both the GB model and \paperacronym{} provide highly accurate energy calculations, with \paperacronym{} achieving comparable performance to the GB model in accuracy, while \paperacronym{}-Lite, although slightly less accurate, still maintains a high level of performance.

Figure~\ref{fig:relative_error_all_system} complements this accuracy analysis by presenting the relative error per atom for each molecule across the different models. This figure plots the relative error in percentage, highlighting how the predicted energy deviates from the true energy on a per-atom basis. The GB model, \paperacronym{}, and \paperacronym{}-Lite are represented by blue squares, green triangles, and red circles, respectively. As observed in the plot, the GB model (blue squares) generally maintains errors close to zero, indicating a consistent performance across different molecules. The \paperacronym{} model (green triangles) also shows a low and stable error distribution, with slightly more variation than the GB model, reflecting its high accuracy as seen in the MAE values. The \paperacronym{}-Lite model (red circles), while still performing well, exhibits a broader range of relative errors, particularly with a few outliers that indicate larger deviations from the true values.

This relative error analysis provides additional insight into the robustness of each model. While \paperacronym{} and \paperacronym{}-Lite both demonstrate strong performance, the GB model consistently achieves the lowest relative errors, affirming its reliability. However, \paperacronym{} and \paperacronym{}-Lite still offer competitive accuracy, with \paperacronym{} performing nearly as well as the GB model, and \paperacronym{}-Lite offering a trade-off between speed and accuracy.

Together, these figures underscore the effectiveness of our \paperacronym{} model in predicting energy values with a high degree of accuracy, while also highlighting areas for potential refinement in the \paperacronym{}-Lite variant.

This description provides a comprehensive analysis of the figures, highlighting key details in both the accuracy comparison and relative error analysis.





\subsection{Model Speed Analysis}
We conducted a comprehensive speed analysis of our models, comparing them against the established PB and GB models using the built-in PB model in AMBER/PBSA as a benchmark. To ensure fairness in the comparison, the force calculation within PBSA was disabled, focusing solely on the time required to calculate the PB energy. 

\begin{figure}[!htbp]
  \centering
  \begin{subfigure}[b]{0.45\textwidth}
    \centering
        \includegraphics[height=0.7\textwidth]{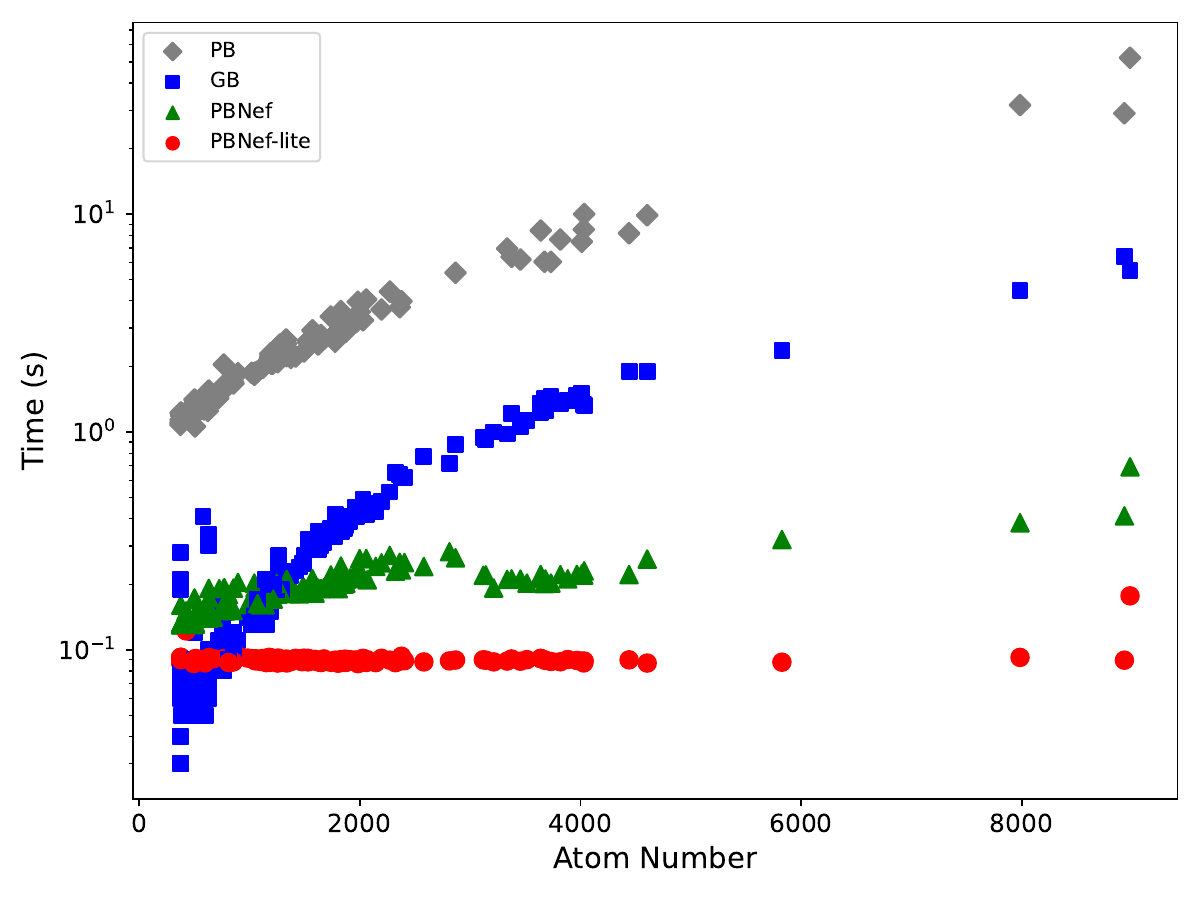}
      \caption{Inference time distribution}
      \label{fig:timing}
  \end{subfigure}
  \begin{subfigure}[b]{0.45\textwidth}
  \centering
        \includegraphics[height=0.7\textwidth]{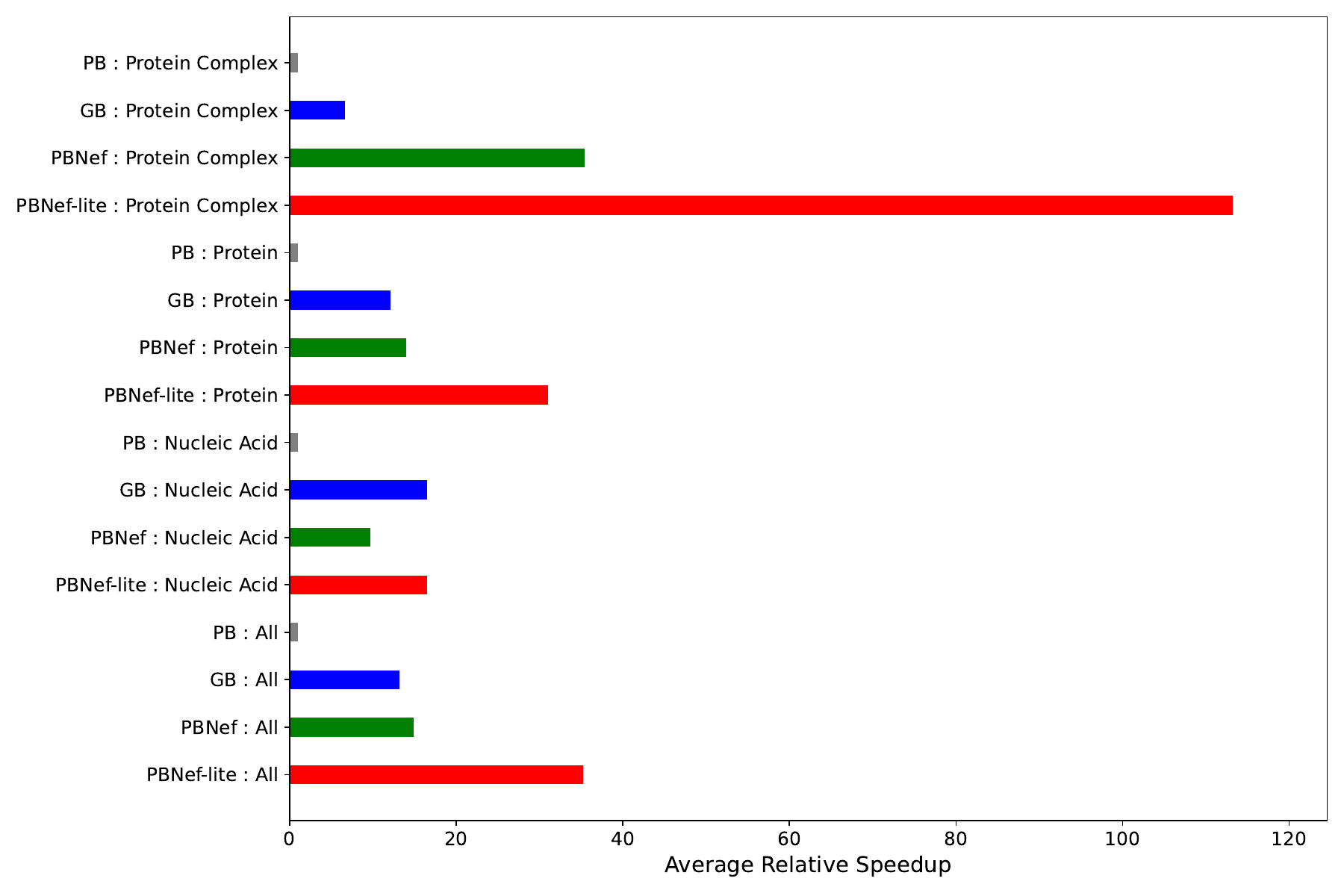}
      \caption{Relative speedup of different methods}
      \label{fig:relative_speed}
  \end{subfigure}
  \caption[Inference Speed Benchmark Results]{
\textbf{Inference Speed Benchmark Results.} 
\textbf{(a) Inference Time Distribution:} This plot illustrates the relationship between the inference time and the number of atoms for four different computational methods: the traditional Poisson-Boltzmann (PB) model (gray diamonds), Generalized Born (GB) model (blue squares), \paperacronym{} (green triangles), and \paperacronym{}-lite (red circles).
\textbf{(b) Relative Speedup of Different Methods:} This bar chart presents the average relative speedup of the GB, \paperacronym{}, and \paperacronym{}-lite methods compared to the PB model across different molecular system categories: proteins, nucleic acids, and protein complexes.
}
  \label{fig:pb_speed}
\end{figure}

Figure~\ref{fig:timing} presents the inference time distribution as a function of the atom count for four different methods: the traditional PB model, the GB model, \paperacronym{}, and \paperacronym{}-lite. As expected, all methods show an increase in calculation time as the number of atoms increases, but the rate of this increase varies significantly among them. The PB model, represented by gray diamonds, shows the steepest increase in computational time with rising atom numbers, highlighting its less favorable scaling. The GB model, represented by blue squares, offers some improvement over PB but still shows significant increases in time with larger systems. In contrast, the \paperacronym{} method, shown as green triangles, exhibits much more favorable scaling characteristics. The increase in computational time is markedly slower compared to both the PB and GB models, particularly as the atom count exceeds 2,000. This trend demonstrates the enhanced efficiency of the \paperacronym{} model in handling larger molecular systems. Even more striking is the performance of \paperacronym{}-lite, depicted by red circles. This variant consistently achieves the fastest inference times across all tested atom numbers, with minimal increases in computation time even as system size grows. This result underscores the robustness and scalability of \paperacronym{}-lite, making it highly suitable for large-scale biomolecular simulations where computational efficiency is critical.

Figure~\ref{fig:relative_speed} further quantifies the efficiency gains by illustrating the relative speedup of the GB, \paperacronym{}, and \paperacronym{}-lite methods compared to the PB model across various molecular systems, including proteins, nucleic acids, and protein complexes. The relative speedup is computed as the ratio of the PB model’s inference time to the inference time of the other methods, averaged over all molecules in each category. The results clearly demonstrate that \paperacronym{}-lite consistently outperforms all other methods, with a particularly pronounced speedup in the protein complex category. Here, \paperacronym{}-lite achieves an average relative speedup exceeding 100-fold compared to the PB model, highlighting its remarkable efficiency in handling highly complex systems. Even in simpler systems like single proteins and nucleic acids, \paperacronym{}-lite achieves speedups of approximately 35-fold and 15-fold, respectively. The \paperacronym{} method also shows significant speedups, achieving up to a 35-fold improvement over the PB model in protein complexes and a 15-fold improvement in proteins, while still maintaining comparable accuracy to the GB model. These results strongly support the conclusion that \paperacronym{} and \paperacronym{}-lite offer substantial computational advantages over traditional methods. The ability of these methods to scale efficiently with system size, while drastically reducing computational time, makes them particularly well-suited for large-scale molecular dynamics simulations. This efficiency is critical for enabling practical simulations of large and complex biomolecular systems, where traditional methods may become computationally prohibitive.



\section{Conclusion}
In conclusion, the proposed \paperacronym{} framework demonstrates significant advancements in the computational modeling of biomolecular electrostatics by efficiently solving the Poisson-Boltzmann equation without the need for explicit partial differential equation solvers. By leveraging neural field representations and advanced data augmentation strategies, \paperacronym{} not only achieves comparable accuracy to traditional methods such as the Generalized Born (GB) and Poisson-Boltzmann (PB) models but also offers substantial improvements in computational speed and scalability. However, it is important to acknowledge the limitations of the \paperacronym{} approach. While \paperacronym{} and its variant \paperacronym{}-Lite exhibit impressive speedups, their accuracy is only comparable to the GB model and does not surpass it. Additionally, these models are not yet capable of predicting the accurate energy contribution of every atom, which limits their precision in detailed electrostatic calculations. We recognize these challenges and will continue to work on improving the accuracy and atom-level precision of \paperacronym{} in future studies.






\clearpage

\bibliography{manuscript}

\begin{thebibliography}{57}
\providecommand{\natexlab}[1]{#1}
\providecommand{\url}[1]{{#1}}
\providecommand{\urlprefix}{URL }
\providecommand{\doi}[1]{\url{https://doi.org/#1}}
\providecommand{\eprint}[2][]{\url{#2}}
 \bibcommenthead

\bibitem[{Achlioptas et~al(2018)Achlioptas, Diamanti, Mitliagkas, and Guibas}]{achlioptas2018learning}
Achlioptas P, Diamanti O, Mitliagkas I, et~al (2018) Learning representations and generative models for 3d point clouds. In: International conference on machine learning, PMLR, pp 40--49

\bibitem[{Baker(2004)}]{baker2004poisson}
Baker NA (2004) Poisson--boltzmann methods for biomolecular electrostatics. In: Methods in enzymology, vol 383. Elsevier, p 94--118

\bibitem[{Baker et~al(2001)Baker, Sept, Joseph, Holst, and McCammon}]{baker2001electrostatics}
Baker NA, Sept D, Joseph S, et~al (2001) Electrostatics of nanosystems: application to microtubules and the ribosome. Proceedings of the National Academy of Sciences 98(18):10037--10041

\bibitem[{Bashford and Case(2000)}]{bashford2000generalized}
Bashford D, Case DA (2000) Generalized born models of macromolecular solvation effects. Annual review of physical chemistry 51(1):129--152

\bibitem[{Batzner et~al(2022)Batzner, Musaelian, Sun, Geiger, Mailoa, Kornbluth, Molinari, Smidt, and Kozinsky}]{batzner20223}
Batzner S, Musaelian A, Sun L, et~al (2022) E (3)-equivariant graph neural networks for data-efficient and accurate interatomic potentials. Nature communications 13(1):2453

\bibitem[{Bruccoleri et~al(1997)Bruccoleri, Novotny, Davis, and Sharp}]{bruccoleri1997finite}
Bruccoleri RE, Novotny J, Davis ME, et~al (1997) Finite difference poisson-boltzmann electrostatic calculations: Increased accuracy achieved by harmonic dielectric smoothing and charge antialiasing. Journal of computational chemistry 18(2):268--276

\bibitem[{Calimet et~al(2001)Calimet, Schaefer, and Simonson}]{calimet2001protein}
Calimet N, Schaefer M, Simonson T (2001) Protein molecular dynamics with the generalized born/ace solvent model. Proteins: Structure, Function, and Bioinformatics 45(2):144--158

\bibitem[{Case et~al(2023)Case, Aktulga, Belfon, Cerutti, Cisneros, Cruzeiro, Forouzesh, Giese, Gootz, Gohlke et~al}]{case2023ambertools}
Case DA, Aktulga HM, Belfon K, et~al (2023) Ambertools. Journal of chemical information and modeling 63(20):6183--6191

\bibitem[{Chen et~al(2007)Chen, Holst, and Xu}]{chen2007finite}
Chen L, Holst MJ, Xu J (2007) The finite element approximation of the nonlinear poisson--boltzmann equation. SIAM journal on numerical analysis 45(6):2298--2320

\bibitem[{Davis and McCammon(1989)}]{davis1989solving}
Davis ME, McCammon JA (1989) Solving the finite difference linearized poisson-boltzmann equation: A comparison of relaxation and conjugate gradient methods. Journal of computational chemistry 10(3):386--391

\bibitem[{Davis and McCammon(1990)}]{davis1990electrostatics}
Davis ME, McCammon JA (1990) Electrostatics in biomolecular structure and dynamics. Chemical Reviews 90(3):509--521

\bibitem[{Davis and McCammon(1991)}]{davis1991dielectric}
Davis ME, McCammon JA (1991) Dielectric boundary smoothing in finite difference solutions of the poisson equation: an approach to improve accuracy and convergence. Journal of computational chemistry 12(7):909--912

\bibitem[{Dominy and Brooks(1999)}]{dominy1999development}
Dominy BN, Brooks CL (1999) Development of a generalized born model parametrization for proteins and nucleic acids. The Journal of Physical Chemistry B 103(18):3765--3773

\bibitem[{Fogolari et~al(2002)Fogolari, Brigo, and Molinari}]{fogolari2002poisson}
Fogolari F, Brigo A, Molinari H (2002) The poisson--boltzmann equation for biomolecular electrostatics: a tool for structural biology. Journal of Molecular Recognition 15(6):377--392

\bibitem[{Gallicchio and Levy(2004)}]{gallicchio2004agbnp}
Gallicchio E, Levy RM (2004) Agbnp: An analytic implicit solvent model suitable for molecular dynamics simulations and high-resolution modeling. Journal of computational chemistry 25(4):479--499

\bibitem[{Gilson et~al(1988)Gilson, Sharp, and Honig}]{gilson1988calculating}
Gilson MK, Sharp KA, Honig BH (1988) Calculating the electrostatic potential of molecules in solution: method and error assessment. Journal of computational chemistry 9(4):327--335

\bibitem[{Grant et~al(2001)Grant, Pickup, and Nicholls}]{grant2001smooth}
Grant JA, Pickup BT, Nicholls A (2001) A smooth permittivity function for poisson--boltzmann solvation methods. Journal of computational chemistry 22(6):608--640

\bibitem[{Hendrycks and Gimpel(2016)}]{hendrycks2016gaussian}
Hendrycks D, Gimpel K (2016) Gaussian error linear units (gelus). arXiv preprint arXiv:160608415

\bibitem[{von Hippel et~al(1984)von Hippel, Bear, Morgan, and McSwiggen}]{von1984protein}
von Hippel PH, Bear DG, Morgan WD, et~al (1984) Protein-nucleic acid interactions in transcription: a molecular analysis. Annual Review of Biochemistry 53(1):389--446

\bibitem[{Holst and Saied(1993)}]{holst1993multigrid}
Holst M, Saied F (1993) Multigrid solution of the poisson—boltzmann equation. Journal of computational chemistry 14(1):105--113

\bibitem[{Holst et~al(2012)Holst, Mccammon, Yu, Zhou, and Zhu}]{holst2012adaptive}
Holst M, Mccammon JA, Yu Z, et~al (2012) Adaptive finite element modeling techniques for the poisson-boltzmann equation. Communications in computational physics 11(1):179--214

\bibitem[{Holst et~al(1994)}]{holst1994poisson}
Holst MJ, et~al (1994) The poisson-boltzmann equation: Analysis and multilevel numerical solution. Applied Mathematics and CRPC, California Institute of Technology

\bibitem[{Honig and Nicholls(1995)}]{honig1995classical}
Honig B, Nicholls A (1995) Classical electrostatics in biology and chemistry. Science 268(5214):1144--1149

\bibitem[{Kraut et~al(2003)Kraut, Carroll, and Herschlag}]{kraut2003challenges}
Kraut DA, Carroll KS, Herschlag D (2003) Challenges in enzyme mechanism and energetics. Annual review of biochemistry 72(1):517--571

\bibitem[{Ladicky et~al(2017)Ladicky, Saurer, Jeong, Maninchedda, and Pollefeys}]{ladicky2017point}
Ladicky L, Saurer O, Jeong S, et~al (2017) From point clouds to mesh using regression. In: Proceedings of the IEEE International Conference on Computer Vision, pp 3893--3902

\bibitem[{Li et~al(2020)Li, Kovachki, Azizzadenesheli, Liu, Bhattacharya, Stuart, and Anandkumar}]{li2020fourier}
Li Z, Kovachki N, Azizzadenesheli K, et~al (2020) Fourier neural operator for parametric partial differential equations. arXiv preprint arXiv:201008895

\bibitem[{Lu et~al(2008)Lu, Zhou, Holst, and McCammon}]{lu2008recent}
Lu B, Zhou Y, Holst M, et~al (2008) Recent progress in numerical methods for the poisson-boltzmann equation in biophysical applications. Commun Comput Phys 3(5):973--1009

\bibitem[{Luo et~al(2002)Luo, David, and Gilson}]{luo2002accelerated}
Luo R, David L, Gilson MK (2002) Accelerated poisson--boltzmann calculations for static and dynamic systems. Journal of computational chemistry 23(13):1244--1253

\bibitem[{Nagelkerke et~al(1991)}]{nagelkerke1991note}
Nagelkerke NJ, et~al (1991) A note on a general definition of the coefficient of determination. biometrika 78(3):691--692

\bibitem[{Nicholls and Honig(1991)}]{nicholls1991rapid}
Nicholls A, Honig B (1991) A rapid finite difference algorithm, utilizing successive over-relaxation to solve the poisson--boltzmann equation. Journal of computational chemistry 12(4):435--445

\bibitem[{Nymeyer and Garcia(2003)}]{nymeyer2003simulation}
Nymeyer H, Garcia AE (2003) Simulation of the folding equilibrium of $\alpha$-helical peptides: a comparison of the generalized born approximation with explicit solvent. Proceedings of the National Academy of Sciences 100(24):13934--13939

\bibitem[{Onufriev and Alexov(2013)}]{onufriev2013protonation}
Onufriev AV, Alexov E (2013) Protonation and pk changes in protein--ligand binding. Quarterly reviews of biophysics 46(2):181--209

\bibitem[{Osher and Fedkiw(2003)}]{osher2003level}
Osher S, Fedkiw R (2003) Level set methods and dynamic implicit surfaces springer. New York Berlin Heidelberg

\bibitem[{Page and Jencks(1971)}]{page1971entropic}
Page MI, Jencks WP (1971) Entropic contributions to rate accelerations in enzymic and intramolecular reactions and the chelate effect. Proceedings of the National Academy of Sciences 68(8):1678--1683

\bibitem[{Pan and Duraisamy(2020)}]{pan2020physics}
Pan S, Duraisamy K (2020) Physics-informed probabilistic learning of linear embeddings of nonlinear dynamics with guaranteed stability. SIAM Journal on Applied Dynamical Systems 19(1):480--509

\bibitem[{Qi et~al(2017)Qi, Su, Mo, and Guibas}]{qi2017pointnet}
Qi CR, Su H, Mo K, et~al (2017) Pointnet: Deep learning on point sets for 3d classification and segmentation. In: Proceedings of the IEEE conference on computer vision and pattern recognition, pp 652--660

\bibitem[{Raissi et~al(2019)Raissi, Perdikaris, and Karniadakis}]{raissi2019physics}
Raissi M, Perdikaris P, Karniadakis GE (2019) Physics-informed neural networks: A deep learning framework for solving forward and inverse problems involving nonlinear partial differential equations. Journal of Computational physics 378:686--707

\bibitem[{Raissi et~al(2020)Raissi, Yazdani, and Karniadakis}]{raissi2020hidden}
Raissi M, Yazdani A, Karniadakis GE (2020) Hidden fluid mechanics: Learning velocity and pressure fields from flow visualizations. Science 367(6481):1026--1030

\bibitem[{Record~Jr et~al(1976)Record~Jr, Lohman, and De~Haseth}]{record1976ion}
Record~Jr MT, Lohman TM, De~Haseth P (1976) Ion effects on ligand-nucleic acid interactions. Journal of molecular biology 107(2):145--158

\bibitem[{Rocchia et~al(2001)Rocchia, Alexov, and Honig}]{rocchia2001extending}
Rocchia W, Alexov E, Honig B (2001) Extending the applicability of the nonlinear poisson- boltzmann equation: multiple dielectric constants and multivalent ions. The Journal of Physical Chemistry B 105(28):6507--6514

\bibitem[{Sharp and Honig(1990)}]{sharp1990electrostatic}
Sharp KA, Honig B (1990) Electrostatic interactions in macromolecules: theory and applications. Annual review of biophysics and biophysical chemistry 19(1):301--332

\bibitem[{Sirignano and Spiliopoulos(2018)}]{sirignano2018dgm}
Sirignano J, Spiliopoulos K (2018) Dgm: A deep learning algorithm for solving partial differential equations. Journal of computational physics 375:1339--1364

\bibitem[{Tancik et~al(2020)Tancik, Srinivasan, Mildenhall, Fridovich-Keil, Raghavan, Singhal, Ramamoorthi, Barron, and Ng}]{tancik2020fourier}
Tancik M, Srinivasan P, Mildenhall B, et~al (2020) Fourier features let networks learn high frequency functions in low dimensional domains. Advances in neural information processing systems 33:7537--7547

\bibitem[{Thuerey et~al(2021)Thuerey, Holl, Mueller, Schnell, Trost, and Um}]{thuerey2021physics}
Thuerey N, Holl P, Mueller M, et~al (2021) Physics-based deep learning. arXiv preprint arXiv:210905237

\bibitem[{Tsui and Case(2000)}]{tsui2000molecular}
Tsui V, Case DA (2000) Molecular dynamics simulations of nucleic acids with a generalized born solvation model. Journal of the American Chemical Society 122(11):2489--2498

\bibitem[{Wang et~al(2000)Wang, Cieplak, and Kollman}]{wang2000well}
Wang J, Cieplak P, Kollman PA (2000) How well does a restrained electrostatic potential (resp) model perform in calculating conformational energies of organic and biological molecules? Journal of computational chemistry 21(12):1049--1074

\bibitem[{Wang et~al(2009)Wang, Cai, Li, Zhao, and Luo}]{wang2009achieving}
Wang J, Cai Q, Li ZL, et~al (2009) Achieving energy conservation in poisson--boltzmann molecular dynamics: Accuracy and precision with finite-difference algorithms. Chemical physics letters 468(4-6):112--118

\bibitem[{Wang and Wade(2003)}]{wang2003implicit}
Wang T, Wade RC (2003) Implicit solvent models for flexible protein--protein docking by molecular dynamics simulation. Proteins: Structure, Function, and Bioinformatics 50(1):158--169

\bibitem[{Warshel and Levitt(1976)}]{warshel1976theoretical}
Warshel A, Levitt M (1976) Theoretical studies of enzymic reactions: dielectric, electrostatic and steric stabilization of the carbonium ion in the reaction of lysozyme. Journal of molecular biology 103(2):227--249

\bibitem[{Warshel and Russell(1984)}]{warshel1984calculations}
Warshel A, Russell ST (1984) Calculations of electrostatic interactions in biological systems and in solutions. Quarterly reviews of biophysics 17(3):283--422

\bibitem[{Wei et~al(2021)Wei, Zhao, and Luo}]{wei2021machine}
Wei H, Zhao Z, Luo R (2021) Machine-learned molecular surface and its application to implicit solvent simulations. Journal of chemical theory and computation 17(10):6214--6224

\bibitem[{Willmott and Matsuura(2005)}]{willmott2005advantages}
Willmott CJ, Matsuura K (2005) Advantages of the mean absolute error (mae) over the root mean square error (rmse) in assessing average model performance. Climate research 30(1):79--82

\bibitem[{Wu et~al(2023)Wu, Wei, Zhu, and Luo}]{wu2023grid}
Wu Y, Wei H, Zhu Q, et~al (2023) Grid-robust efficient neural interface model for universal molecule surface construction from point clouds. The Journal of Physical Chemistry Letters 14(40):9034--9041

\bibitem[{Ye et~al(2010)Ye, Wang, and Luo}]{ye2010revised}
Ye X, Wang J, Luo R (2010) A revised density function for molecular surface calculation in continuum solvent models. Journal of chemical theory and computation 6(4):1157--1169

\bibitem[{Zakrzewska et~al(1996)Zakrzewska, Madami, and Lavery}]{zakrzewska1996poisson}
Zakrzewska K, Madami A, Lavery R (1996) Poisson-boltzmann calculations for nucleic acids and nucleic acids complexes. Chemical physics 204(2-3):263--269

\bibitem[{Zhang et~al(2018)Zhang, Han, Wang, Car, and E}]{zhang2018deep}
Zhang L, Han J, Wang H, et~al (2018) Deep potential molecular dynamics: a scalable model with the accuracy of quantum mechanics. Physical review letters 120(14):143001

\bibitem[{Zhu and Zabaras(2018)}]{zhu2018bayesian}
Zhu Y, Zabaras N (2018) Bayesian deep convolutional encoder--decoder networks for surrogate modeling and uncertainty quantification. Journal of Computational Physics 366:415--447

\end{thebibliography}

\end{document}